\documentclass[a4paper,11pt]{article}
\usepackage{amsmath,amsfonts,amssymb,times,mathptmx,graphicx,amsthm,lineno}
\usepackage{placeins}
\usepackage[T1]{fontenc}
\usepackage[latin1]{inputenc}
\title{\bf Positioning Error Probability for Some Forms of
Center-of-Gravity Algorithms Calculated with the Cumulative Distributions. Part I.
 }
\author{Gregorio Landi$^a$\thanks{Corresponding
author. Gregorio.Landi@fi.infn.it}~,   Giovanni E. Landi$^b$\\
\\
\llap{$^a$} Dipartimento di Fisica e Astronomia,
Universita' di Firenze and INFN\\
Largo E. Fermi 2 (Arcetri) 50125, Firenze, Italy\\
\\
\llap{$^b$} ArchonVR S.a.g.l.,\\
Via Cisieri 3,
6900 Lugano, Switzerland.\\ \\
{ June 1, 2020}}
\date{ }
\begin{document}
\maketitle 
\begin{abstract}
To complete a previous paper, the probability density functions of
the center-of-gravity as positioning algorithm
are derived with classical methods. These methods, as
suggested by the textbook of Probability, require
the preliminary calculation of the cumulative distribution functions.
They are more complicated than those previously used
for these tasks. In any case, the cumulative probability distributions
could be useful. The combinations of random
variables are those essential for track fitting  $x={\xi}/{(\xi+\eta)}$,
$x=\theta(x_3-x_1) (-x_3)/(x_3+x_2) +\theta(x_1-x_3)x_1/(x_1+x_2)$
and $x=(x_1-x_3)/(x_1+x_2+x_3)$. The first combination is a partial form of the
two strip center-of-gravity. The second is the complete form, and the third is a
simplified form of the three strip center-of-gravity.
The cumulative probability distribution of
the first expression was reported in the previous publications. The standard
assumption is that $\xi$, $\eta$, $x_1$, $x_2$ and $x_3$ are independent
random variables.
\end{abstract}



\tableofcontents

\pagenumbering{arabic} \oddsidemargin 0cm  \evensidemargin 0cm


\section{Introduction}

In reference~\cite{landi10}, the probability density functions (PDFs) of
positioning errors for some forms of Center-of-Gravity (COG) algorithms were
reported. Their derivations were obtained with a straightforward methods, the
Fermi golden rule $\#1$, an usual tool to handle hard constraints in quantum mechanics.
Of those PDFs, only one was derived with the standard textbook
method (for example~\cite{gnedenko}) through the calculation of
the cumulative probability function. This type of
derivation was heavily synthesized in an appendix of ref.~\cite{landi10}.
Here, our aim is to use the textbook-style calculation for the other two most important
COG algorithms: the complete two-strip COG and the simple three-strip COG.
In any reference to these PDFs, we stated that our first developments were
with the calculations of the cumulative probability and its successive
differentiation, thus, we report them in the following.
We never used the cumulative probability functions
in our track reconstructions, but we can not exclude
their possible utility.
However, we encountered a hard animosity against our approaches
that suggests us to be redundant to avoid criticisms.
For example, to counteract the criticisms of our readers we had to produce two
different demonstrations (refs.~\cite{landi08,landi09}) about the
superiority of our fitting methods compared the standard ones.
We supposed, as usual, that a set of simulations suffices. Instead
very complex and length demonstrations were pretended as {\em condition sine qua non}
for the publication of ref.~\cite{landi07}. We rejected this pretense
considering that as substantial distortion of ref.~\cite{landi07}, without no
additional contribution to our results.  In
our plans, ref.~\cite{landi07} was a phenomenological discussion of two
of our unexpected results, the linear growth and the lucky model, without
resorting to long equations.
In any case, refs.~\cite{landi08,landi09} contain all the equations missed
to ref.~\cite{landi07} (for the referee point of view).
The PDFs, derived here and in ref.~\cite{landi10}, are one of the key elements of our
approach of refs.~\cite{landi05,landi06}.
The other key element is the theorem of ref.~\cite{landi05},
this is essential to insert the functional dependence from
the impact point in the PDFs. In fact, it is the
estimation of the track impact points the aim of all these
developments. Figures 15 and 16 of ref.~\cite{landi05} illustrate
very well the ability of these methods to estimate the track impact points and the
unexpected importance of the Cauchy-(Agnesi) tails.
A detailed discussion of the use of the theorem of ref.~\cite{landi05} to complete
the PDFs with the functions of the track impact points will be the subject of a
future paper.
When we started this study of the COG PDFs,
we expected that a large part of these developments were well known, the COG
algorithms are in use by a long time. With our surprise, we discovered that nobody
worried to calculate them. The belief that all the probabilities are Gaussian
functions is very hard to die.
The perception to move in an unexplored land obliged us to pay attention to any
detail and to select the best allowed path. This strategy was very rewarding,
producing good expected outcomes and excellent unexpected results.
The following are our original {\tt LateX}-notes where we suppressed the parts reported
in ref.~\cite{landi10}. The last subsection contains figures of the PDFs for three strip
COG compared with the PDFs for two strip COG.

\section{Simple two strip case}

\begin{figure} [h!]
\begin{center}
\includegraphics[scale=0.6]{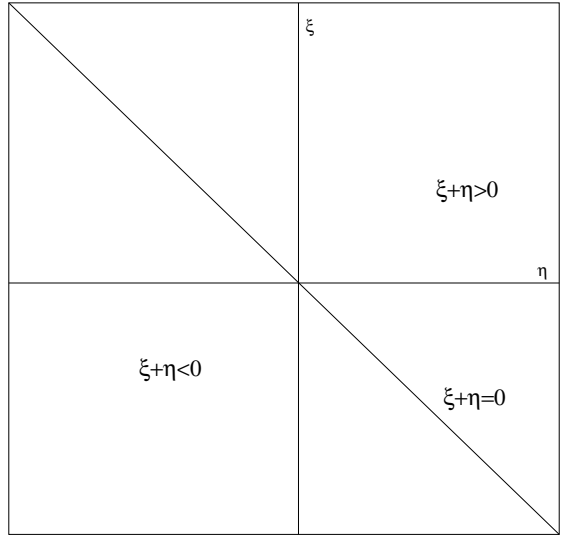}
\caption{\em Sector of the plane $(\eta,\xi)$ where $\xi+\eta>0$ or
$\xi+\eta<0$ and its boundary $\xi+\eta=0$ }\label{fig:plot_1}
\end{center}
\end{figure}

Now we calculate the cumulative probability of a simple two-strip COG,
extending the compressed discussion of ref~\cite{landi10}.
The parameters of the problem are the two energies collected by adjacent strips,
each one affected by an additive random noise (probably Gaussian as our
data look to support). We will proceeds along the lines of the application
of ref.~\cite{gnedenko} to a ratio of two random variables. In our case
we have two random variables but the ratio is that of
a COG $\xi/(\xi+\eta)$. As in ref.~\cite{gnedenko} we will consider the regions where:
\begin{equation}\label{eq:equation_1}
    F_2(x)=P(\frac{\xi}{\xi+\eta}\leq x)
\end{equation}
The derivative of Eq.~\ref{eq:equation_1} respect to x gives the PDF of this case.
We have first two conditions:
\begin{itemize}
   \item $\xi+\eta\geq 0$
   \item $\xi+\eta < 0$
\end{itemize}
The plane $(\eta,\xi)$ is divided in two parts by the line $\xi=-\eta$ (figure~\ref{fig:plot_1}).

\begin{figure} [h!]
\begin{center}
\includegraphics[scale=0.6]{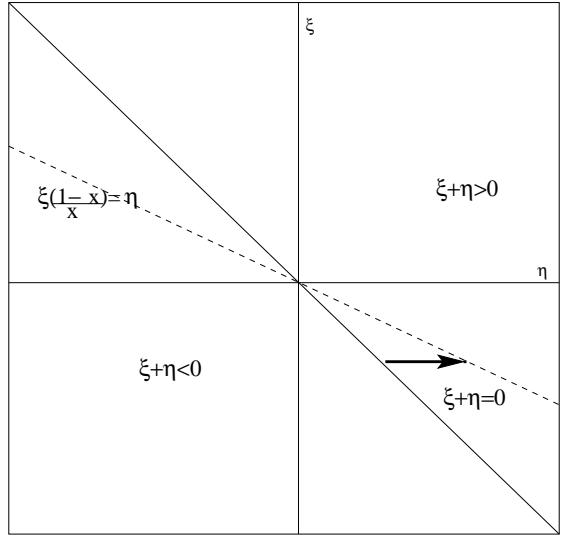}
\caption{\em Sector of the plane $(\eta,\xi)$ where $\xi+\eta>0$ or
$\xi+\eta<0$ and its boundary $\xi+\eta=0$, the dashed line is the line $\xi(1-x)/x=\eta$ for negative
$x$. The arrow indicates the integration path for $(\xi+\eta)>0$ }\label{fig:plot_2}
\end{center}
\end{figure}

 We have various conditions to satisfy to find $F_2(x)$.

\begin{equation}
\begin{aligned}
&\frac{\xi}{\xi+\eta}<x \ \ \rightarrow \xi+\eta>0 \Rightarrow \xi<x(\xi+\eta)\ \ \ \Rightarrow \
\xi(1-x)<x\eta\\
&x<0 \ \  \Rightarrow \ \xi\frac{(1-x)}{x}>\eta \ \ \ \ \ \ \ \ \ x>0 \ \ \Rightarrow \ \
\xi\frac{(1-x)}{x}<\eta
\end{aligned}
\end{equation}
 and similarly we have:
\begin{equation}
\begin{aligned}
&\frac{\xi}{\xi+\eta}<x \ \ \rightarrow \xi+\eta<0 \Rightarrow \xi>x(\xi+\eta)\ \ \ \Rightarrow \
 \xi(1-x)>x\eta\\
    &x<0 \ \  \Rightarrow \ \xi\frac{(1-x)}{x}<\eta \ \ \ \ \ \ \ \ \ x>0 \ \ \Rightarrow \ \
    \xi\frac{(1-x)}{x}>\eta
\end{aligned}
\end{equation}
From the definition of $x$, we see that the equation $\xi=0$ is obtained with $x=0$. For
$x<0$, the function $\xi(1-x)/x=\eta$ is the dashed line  reported in fig.~\ref{fig:plot_2}. All
the lines for $x<0$ go trough the origin with the slope converging to the $\eta$-axis for $x\rightarrow 0$.

\begin{figure} [h]
\begin{center}
\includegraphics[scale=0.6]{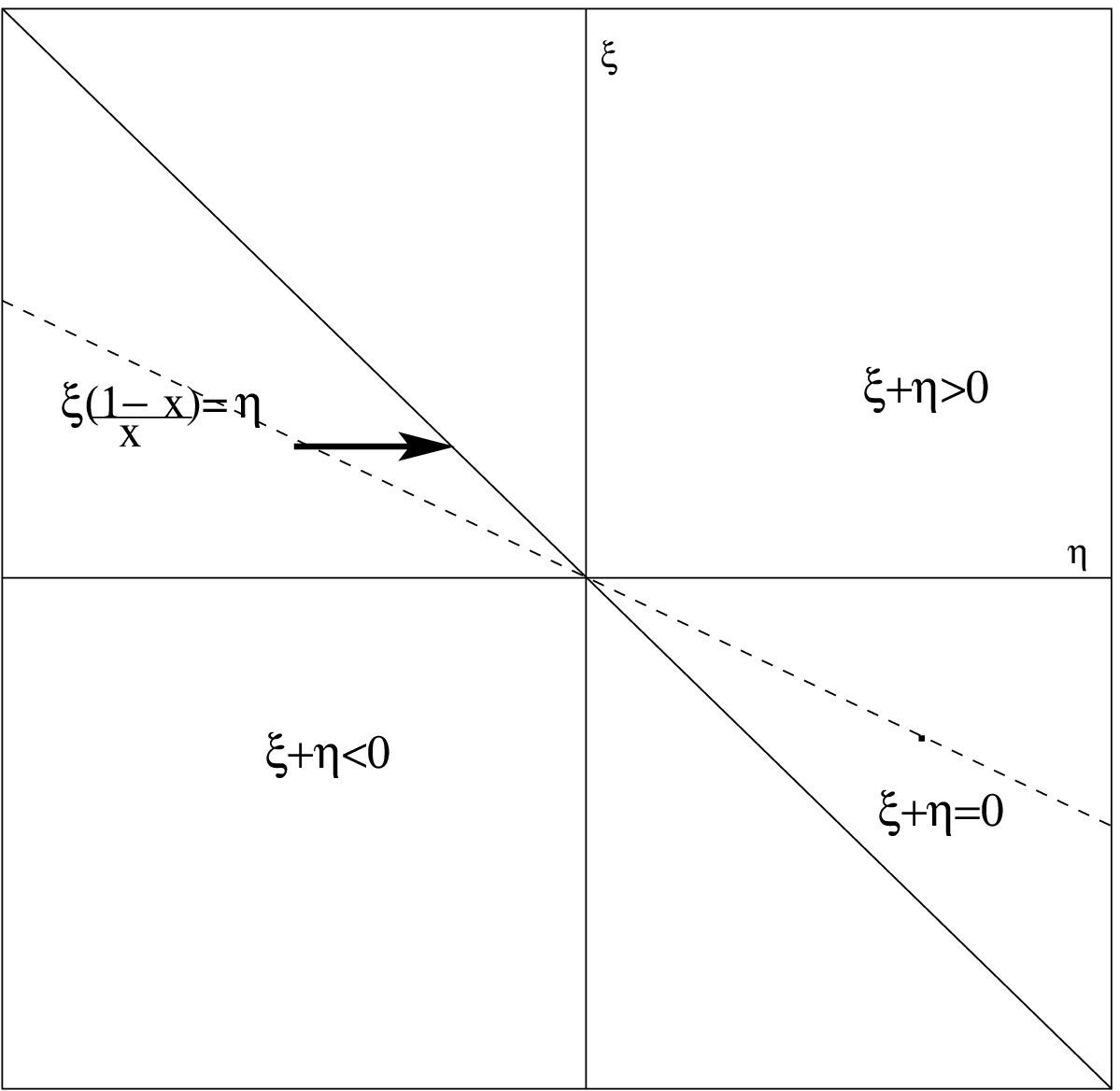}
\caption{\em Sector of the plane $(\eta,\xi)$ where $\xi+\eta>0$ or
$\xi+\eta<0$ and its boundary $\xi+\eta=0$, the dashed line is the line $\xi(1-x)/x=\eta$ for negative $x$, the arrow indicates the $\eta$ integration-path for $(\xi+\eta)<0$ }\label{fig:figure_3}
\end{center}
\end{figure}

Given an $x<0$ and $\xi+\eta>0$,  the integration region is that in the lower half plane within the two
lines $\xi+\eta=0$ and the dashed line. The integration directions must give a positive area, so a part
of the integral on $\eta$ is indicated in fig.~\ref{fig:plot_2} as a thick arrow. The contribution to
$F_2(x)$ of this part of the plane is $F_2^a(x)$:
\begin{equation}
    F_2^a(x)=\int_{-\infty}^0\mathrm{d}\xi P_1(\xi)\int_{-\xi}^{\xi(1-x)/x} P_2(\eta)\mathrm{d}\eta.
\end{equation}
$P_1(\xi)$ and $P_2(\eta)$ are the two distributions of the random variables $\xi$ and $\eta$. The
contribution to $F_2(x)$ from the region with $(\xi+\eta)<0$ is given by the upper part of plane within
the two lines $\xi(1-x)/x=\eta$ and $\xi=-\eta$. A part of the integration path in $\eta$ is indicated
with a thick arrow in fig.~\ref{fig:figure_3}.

Its contribution to $F_2(x)$, we call it $F_2^b(X)$, is given by:
\begin{equation}
 F_2^b(x)=\int_0^{\infty}\mathrm{d}\xi P_1(\xi)\int_{\xi(1-x)/x}^{-\xi} P_2(\eta)\mathrm{d}\eta.
\end{equation}

\begin{figure} [h!]
\begin{center}
\includegraphics[scale=0.6]{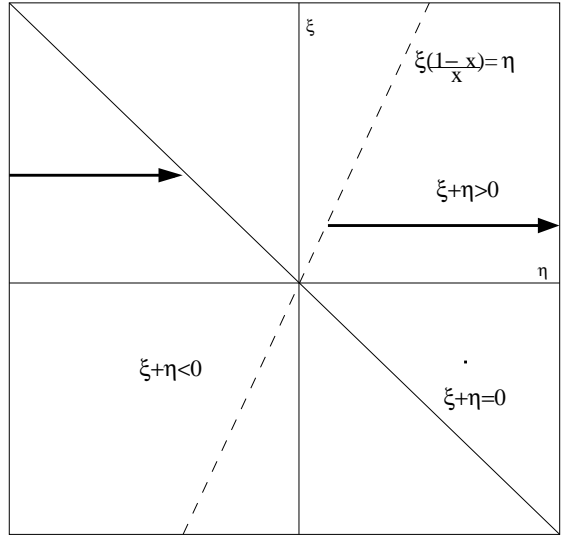}
\caption{\em Sector of the plane $(\eta,\xi)$ where $\xi+\eta>0$ or
$\xi+\eta<0$ and its boundary $\xi+\eta=0$, the dashed line is the line $\xi(1-x)/x=\eta$ for
positive $x$. The integration regions are that with the arrows and must cover positive and negative
values of $\xi$  }\label{fig:figure_4}
\end{center}
\end{figure}

For $x>0$ the lines are typically that reported in  figure~\ref{fig:figure_4}. Now we have to integrate on a larger sector. An integration must cover the region up to the line with $x=0$, (the $\eta$-axis), and an integration must cover the remaining part up to the line $\xi(1-x)/x=\eta$ for each $(\xi+\eta)>0$ and $(\xi+\eta)<0$. $F_2^c(x)$ and $F_2^d(x)$ respectively for $(\xi+\eta)>0$ and $(\xi+\eta)<0$ are given by:
\begin{equation}
F_2^c(x)=\int_{-\infty}^0\mathrm{d}\xi P_1(\xi)\int_{-\xi}^{+\infty}P_2(\eta)\mathrm{d}\eta+\int_0^{+\infty}\mathrm{d}\xi\,P_1(\xi)\int_{\xi(1-x)/x}^{+\infty}P_2(\eta)\mathrm{d}\eta
\end{equation}
\begin{equation*}
F_2^d(x)=\int_{-\infty}^0\mathrm{d}\xi P_1(\xi)\int_{-\infty}^{\xi(1-x)/x}P_2(\eta)\mathrm{d}\eta+\int_0^{+\infty}\mathrm{d}\xi\,P_1(\xi)\int_{-\infty}^{-\xi}P_2(\eta)\mathrm{d}\eta
\end{equation*}
So for $x\leq 0$ we obtain:
\begin{equation}\label{eq:equation_5}
   F_2(x)=\int_{-\infty}^0\mathrm{d}\xi P_1(\xi)\int_{-\xi}^{\xi(1-x)/x} P_2(\eta)\mathrm{d}\eta+\int_0^{\infty}\mathrm{d}\xi P_1(\xi)\int_{\xi(1-x)/x}^{-\xi} P_2(\eta)\mathrm{d}\eta
\end{equation}
and $x>0$ $F_2(x)$ is:
\begin{equation}
\begin{aligned}
F_2(x)=&\int_{-\infty}^0\mathrm{d}\xi
 P_1(\xi)\int_{-\xi}^{+\infty}P_2(\eta)\mathrm{d}\eta+\int_0^{+\infty}\mathrm{d}\xi\,P_1(\xi)\int_{\xi(1-x)/x}^{+\infty}
P_2(\eta)\mathrm{d}\eta+\\
&\int_{-\infty}^0\mathrm{d}\xi
P_1(\xi)\int_{-\infty}^{\xi(1-x)/x}P_2(\eta)\mathrm{d}\eta+\int_0^{+\infty}\mathrm{d}\xi\,P_1(\xi)\int_{-\infty}^{-\xi}
P_2(\eta)\mathrm{d}\eta
\end{aligned}
\end{equation}
It is easy to verify the consistency of $F_2(x)$, in fact $F_2(x\rightarrow-\infty)=0$ and
 $F_2(x\rightarrow+\infty)=1$. When $x\rightarrow-\infty$ it is $(1-x)/x\rightarrow-1$ and:
\begin{equation}
    F_2(x\rightarrow-\infty)=\int_{-\infty}^0\mathrm{d}\xi P_1(\xi)\int_{-\xi}^{-\xi} P_2(\eta)\mathrm{d}\eta+\int_0^{\infty}\mathrm{d}\xi P_1(\xi)\int_{-\xi}^{-\xi} P_2(\eta)\mathrm{d}\eta=0
\end{equation}
the integration on $\eta$ gives zero in the two integrals.
For $x\rightarrow+\infty$ and $(1-x)/x\rightarrow-1$ it is:
\begin{equation}
\begin{aligned}
F_2(x)=&\int_{-\infty}^0\mathrm{d}\xi P_1(\xi)\int_{-\xi}^{+\infty}P_2(\eta)\mathrm{d}\eta+\int_0^{+\infty}\mathrm{d}\xi\,P_1(\xi)\int_{-\xi}^{+\infty}P_2(\eta)\mathrm{d}\eta+\\
&\int_{-\infty}^0\mathrm{d}\xi P_1(\xi)\int_{-\infty}^{-\xi}P_2(\eta)\mathrm{d}\eta+\int_0^{+\infty}\mathrm{d}\xi\,P_1(\xi)\int_{-\infty}^{-\xi}P_2(\eta)\mathrm{d}\eta\\
=&\int_{-\infty}^{+\infty}P_2(\xi)\mathrm{d}\xi\int_{-\infty}^{+\infty}P_1(\eta)\mathrm{d}\eta=1
\end{aligned}
\end{equation}
that is equal to one for the normalization of $P_1$ and $P_2$.
This check assures the absence of trivial errors.

\subsection{Probability Density Function}

The probability density function (PDF) is extracted by the function $F_2(x)$ with a derivative
 respect to $x$. Due to the complex dependence from $x$ of the boundaries of integration, the
 derivative must be done with the definition of an auxiliary variable $y(x)$ and multiplying the
 derivative respect to $y$  for $\mathrm{d} y/\mathrm{d} x$. We will indicate this PDF with $P_{xg2R}(x)$.
 The result is:

\begin{equation}\label{eq:equation_10}
    P_{xg2R}(x)=\frac{\mathrm{d} F_2(x)}{\mathrm{d} x}=\frac{1}{x^2}\big [\int_0^{+\infty}\mathrm{d}\xi\,P_1(\xi)\xi
    P_2(\xi\frac{1-x}{x})-
    \int_{-\infty}^0\,\mathrm{d}\xi\,P_1(\xi)\xi\,P_2(\xi\frac{1-x}{x})\big].
\end{equation}

An identical result is obtained differentiating $F_2(x)$ for $x\geq 0$ or $F_2(x)$ for $x<0$.
Equation~\ref{eq:equation_10} is the PDF of the two strip COG assuming the independence of the
two strip noise and the strip $1$ is the right strip. The factor $1/x^2$ and the factor $\xi$
in the integral come from the derivative of $\xi(1-x)/x$ respect to $x$.

To consider the left strip one can proceed
as for the right strip obtaining different figures. The final result is equivalent to substitute
to the probability $P_1$ of the right strip the probability $P_3$ of the left strip and change $x$
in $-y$ to obtain the PDF $P_{xg2L}(y)$ of the two strip COG with the left strip.

\subsection{Cumulative probability using the left strip}

Even if the steps to obtain this distribution are identical to the previous ones, with a small
modifications we will reproduce the path because it will be used in the following.
\begin{equation}
 y=\frac{-\beta}{\beta+\eta}
\end{equation}

\begin{figure} [h!]
\begin{center}
\includegraphics[scale=0.6]{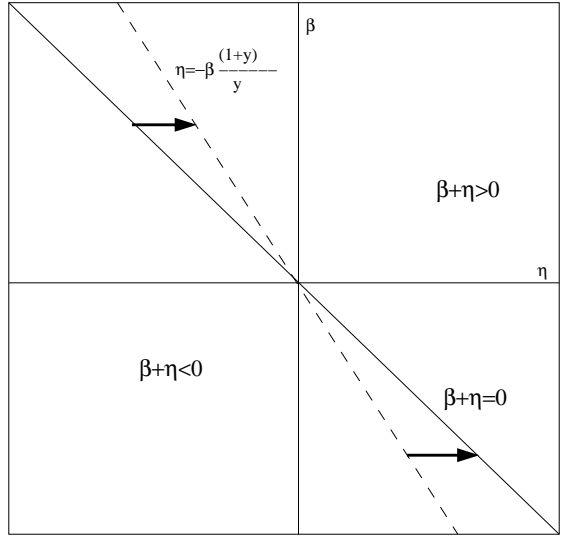}
\caption{\em Integration regions for $y<0$ for the two strip COG with the left
 strip}\label{fig:figure_5a}
\end{center}
\end{figure}

As usual we calculate the function $F_{2L}(y)$ defined as the region where
$-\beta/(\beta+\eta)<y$. We have to separate the two regions $\beta+\eta\geq 0$ and $\beta+\eta<0$
Now the integrals for $y<0$ has to be done as in fig.\ref{fig:figure_5a}

The function $F_{2L}(y)$ for $y<0$ is given by:
\begin{equation}
 F_{2L}(y)=\int_0^{\infty}\mathrm{d}\beta P_3(\beta)\int_{-\beta}^{\beta(-1-y)/y}
  P_2(\eta)\mathrm{d}\eta+\int_{-\infty}^0\mathrm{d}\beta P_3(\beta)\int_{\beta(-1-y)/y}^{-\beta} P_2(\eta)\mathrm{d}\eta
\end{equation}

\begin{figure} [h!]
\begin{center}
\includegraphics[scale=0.6]{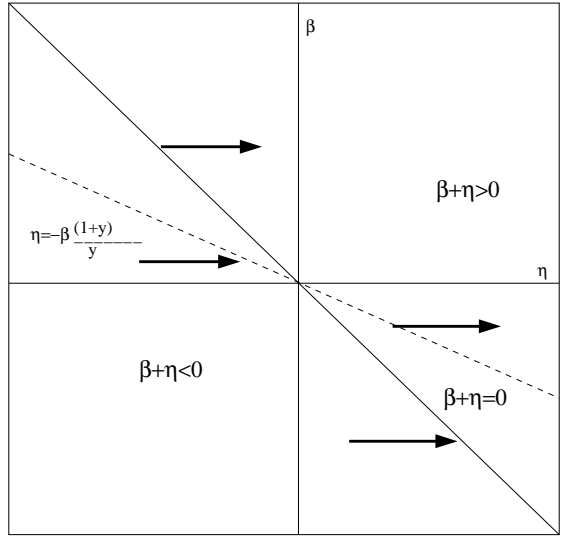}
\caption{\em Integration regions for $y<0$ for the two strip COG with the left
 strip}\label{fig:figure_5b}
\end{center}
\end{figure}

For $y\geq 0$ the integration regions are that of fig.~\ref{fig:figure_5b}, obviously even above
and below the $\beta=0$ line. The function $F_{2L}(y)$ for $y\geq 0$ is given by:
\begin{equation}
\begin{aligned}
    F_{2L}(y)=&\int_0^{+\infty}\,\mathrm{d}\beta\,P_3(\beta)\int_{-\beta}^{+\infty}P_2(\eta)\mathrm{d}\eta +
    \int_{-\infty}^0\,\mathrm{d}\beta\,P_3(\beta)\int_{\beta(-1-y)/y}^{+\infty}\,P_2(\eta)\,\mathrm{d}\eta+\\
    &\int_{-\infty}^0\mathrm{d}\beta\,P_3(\beta)\int_{-\infty}^{-\beta}\,P_2(\eta)\,\mathrm{d}\eta+
    \int_0^{+\infty}\,\mathrm{d}\beta\,P_3(\beta)\int_{-\infty}^{\beta(-1-y)/y}\,P_2(\eta)\mathrm{d}\eta\\
\end{aligned}
\end{equation}

To control the consistency of the equations, the limit of $y\rightarrow\infty$ must give one. It
is easy to verify this as in the approach of $F_{2R}(x)$.

Differentiating $F_{2L}(y)$ respect to $y$ gives the PDF for the COG distribution with
the left strip.
\begin{equation}\label{eq:equation_20}
  P_{xg2L}(y)=\frac{1}{y^2}\big[\int_0^{+\infty}\mathrm{d}\beta\,P_3(\beta)\beta\,P_2(\frac{-1-y}{y}\eta)-
  \int_{-\infty}^0\mathrm{d}\beta P_3(\beta)\beta\,P_2(\frac{-1-y}{y}\beta) \big].
\end{equation}

The variables $\beta,\xi,\eta$ can be approximated as an average value and an additive Gaussian
noise. In this assumption the PDF $P_1,P_2,P_3$ can be represented as Gaussian functions with
 averages corresponding to the unperturbed energy values collected by the strips.
With MATHEMATICA~\cite{MATHEMATICA}, it is possible to obtain an exact
form of the above integrals with Gaussian PDF. This full form is very complex,
but we have to consider that we are interested
in $x$ or $y$ values less than or equal to $0.5$ in absolute value.

The full form of the PDF has a very unusual behavior for $x\rightarrow\infty$. It goes to $\infty$
as $constant/x^2$ with an extremely small constant ($10^{-40}$). This convergence to $\infty$ is very
slow compared to a Gaussian function.  The PDF has infinite variance and
no average. Thus, for these PDF the central limit theorem cannot be applied. Numerical calculations
does not reveal the divergence.

\section{The complete two strip center-of-gravity}

\begin{figure} [h!]
\begin{center}
\includegraphics[scale=0.6]{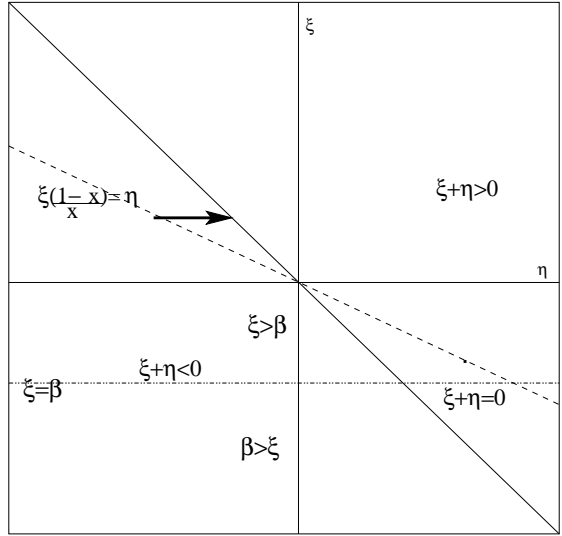}
\caption{\em This is fig.~\ref{fig:figure_3} with the dash-dotted line indicating
the boundary of the
region with $\xi\geq\beta$. Now the integration region is above this line, here we have $x<0$.
}\label{fig:figure_3a}
\end{center}
\end{figure}

The equations developed up to now consider only two strips. The full two strip algorithm selects
the second strip as the greatest of two (left and right) nearby strips. For $x_g\approx 0$, the
noise can favor one or the other strip. So, the energy of the third strip can plays its role
in generating the typical anomaly of the two strip COG. The addition of the third strip requires
to study the function $F_2(x)$ in three dimensions. The combinations of the previous plots will
be useful in this case. As in the previous sections, the three strips are indicated with the
index $1$ for the right strip, with the index $2$ the central strip, and the index $3$ with
the left strip. The random variables are $\xi$, $\eta$ and $\beta$ for the right, central and left
strip and their axis are directed as $y$, $x$ and $z$ axis of a 3D reference system . If the
energy of the right strip is much higher than that of left strip, the effect of
the left strip is negligible. At $a_3\rightarrow-\infty$ the calculation of $F_2(x)$ is identical
to that of fig.~\ref{fig:plot_2},~\ref{fig:figure_3},~\ref{fig:figure_4}. When $\xi\geq\beta$ one
get eq.~\ref{eq:equation_5} and multiplies $F_2$ by $P_3(\beta)$ and limits the integrations to
the due regions $\xi\geq\beta$ as plotted in fig.~\ref{fig:figure_3a}.

\begin{figure} [h!]
\begin{center}
\includegraphics[scale=0.6]{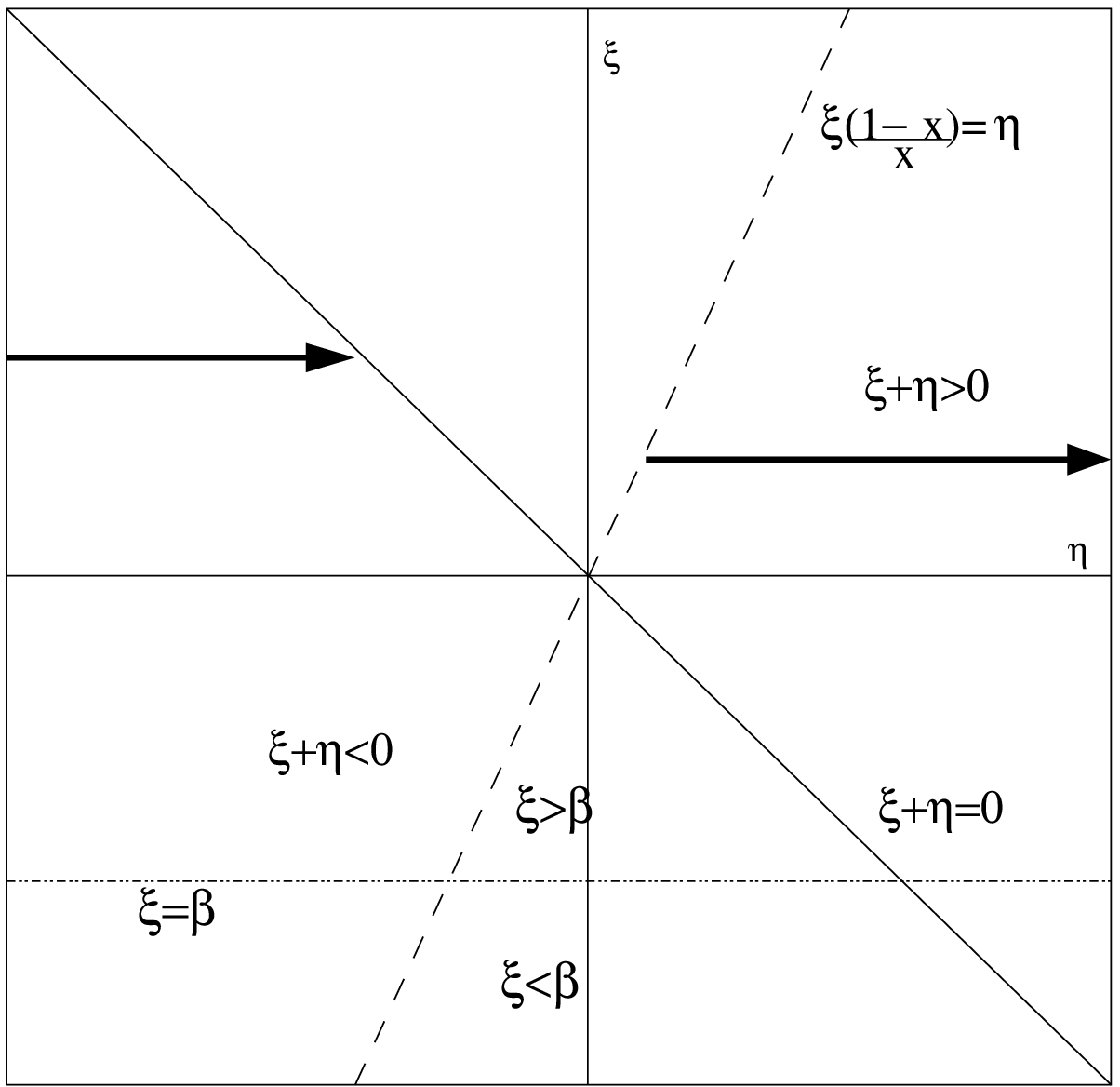}
\caption{\em This is fig.~\ref{fig:figure_4} with the dash-dotted line indicating
the boundary of the
region with $\xi\geq\beta$. The integration region is above this line, here we have $x>0$.
}\label{fig:figure_4a}
\end{center}
\end{figure}

When $\beta>\xi$ the function $F_2(y)$ is used
with the substitution $y\rightarrow x$ and $P_1(\xi)\rightarrow P_3(\beta)$ and integrating over
the regions $\beta\geq\xi$. The result is:

\[x<0\]

\begin{equation}\label{eq:equation_40}
\begin{aligned}
F_3(x)=&\int_{-\infty}^0\,\mathrm{d}\beta\,P_3(\beta)\int_{\beta}^0\mathrm{d}\xi P_1(\xi)\int_{-\xi}^{\xi(1-x)/x}
 P_2(\eta)\mathrm{d}\eta+\\
&\int_0^\infty\,\mathrm{d}\beta\,P_3(\beta)\,\int_\beta^{\infty}\mathrm{d}\xi P_1(\xi)\int_{\xi(1-x)/x}^{-\xi}
 P_2(\eta)\mathrm{d}\eta+\\
&\int_{-\infty}^0\,\mathrm{d}\beta\,P_3(\beta)\int_0^\infty\mathrm{d}\xi P_1(\xi)\int_{\xi(1-x)/x}^{-\xi}
 P_2(\eta)\mathrm{d}\eta+\\
 &\int_{-\infty}^0\,\mathrm{d}\xi\,P_1(\xi)\,\int_{\xi}^0\mathrm{d}\beta P_3(\beta)\int_{\beta(-1-x)/x}^{-\beta}
  P_2(\eta)\mathrm{d}\eta+\\
 &\int_0^{\infty}\,\mathrm{d}\xi\,P_1(\xi)\,\int_\xi^{\infty}\mathrm{d}\beta P_3(\beta)\int_{-\beta}^{\beta(-1-x)/x}
  P_2(\eta)\mathrm{d}\eta\\
&\int_{-\infty}^0\,\mathrm{d}\xi\,P_1(\xi)\,\int_0^{\infty}\mathrm{d}\beta P_3(\beta)\int_{-\beta}^{\beta(-1-x)/x}
  P_2(\eta)\mathrm{d}\eta\\
\end{aligned}
\end{equation}
The condition $\lim_{x\rightarrow-\infty}F_3(x)=0$ is easily verified due to the coincidence of
the two integration limits of the inner integrals.
The integration regions for $x\geq 0$ are illustrated in fig.~\ref{fig:figure_4a}.

\[x\geq 0\]

\begin{equation}\label{eq:equation_50}
\begin{aligned}
&F_3(x)=\\
&\int_{-\infty}^0\,\mathrm{d}\beta
P_3(\beta)\int_{\beta}^0\mathrm{d}\xi P_1(\xi)\int_{-\xi}^{+\infty}P_2(\eta)\mathrm{d}\eta+
\int_0^\infty\mathrm{d}\beta P_3(\beta)\int_\beta^{+\infty}\mathrm{d}\xi P_1(\xi)\int_{\xi(1-x)/x}^{+\infty}
P_2(\eta)\mathrm{d}\eta+\\
&\int_{-\infty}^0\mathrm{d}\beta P_3(\beta)\int_{\beta}^0\mathrm{d}\xi
P_1(\xi)\int_{-\infty}^{\xi(1-x)/x}P_2(\eta)\mathrm{d}\eta+\int_0^\infty\mathrm{d}\beta
 P_3(\beta)\int_\beta^{+\infty}P_1(\xi)\mathrm{d}\xi\int_{-\infty}^{-\xi}
P_2(\eta)\mathrm{d}\eta+\\
&\int_0^\infty\mathrm{d}\xi P_1(\xi)\int_\xi^{+\infty}\mathrm{d}\beta P_3(\beta)\int_{-\beta}^{+\infty}\mathrm{d}\eta
P_2(\eta) +\int_{-\infty}^0\mathrm{d}\xi P_1(\xi)\int_{\xi}^0\mathrm{d}\beta
 P_3(\beta)\int_{\beta(-1-x)/x}^{+\infty}\mathrm{d}\eta P_2(\eta)+\\
    &\int_{-\infty}^0\mathrm{d}\xi P_1(\xi)\int_{\xi}^0\mathrm{d}\beta
    P_3(\beta)\int_{-\infty}^{-\beta}P_2(\eta)\mathrm{d}\eta+
    \int_0^\infty\mathrm{d}\xi P_1(\xi)\int_\xi^{+\infty}\mathrm{d}\beta
     P_3(\beta)\int_{-\infty}^{\beta(-1-x)/x}P_2(\eta)\mathrm{d}\eta+\\
&\int_{-\infty}^0\mathrm{d}\beta P_3(\beta)\int_0^\infty\mathrm{d}\xi P_1(\xi)\int_{\xi(1-x)/x}^\infty\mathrm{d}\eta
P_2(\eta)+\int_{-\infty}^0\mathrm{d}\beta P_3(\beta)\int_0^\infty\mathrm{d}\xi P_1(\xi)\int_{-\infty}^{-\xi}\mathrm{d}\eta
 P_2(\eta)+\\
&\int_{-\infty}^0\mathrm{d}\xi P_1(\xi)\int_0^\infty\mathrm{d}\beta P_3(\beta)\int_{-\beta}^\infty\mathrm{d}\eta P_2(\eta)+
\int_{-\infty}^0\mathrm{d}\xi P_1(\xi)\int_0^\infty\mathrm{d}\beta P_3(\beta)\int_{-\infty}^{\beta(-1-x)/x}\mathrm{d}\eta
P_2(\eta)
\end{aligned}
\end{equation}

This complex set of integrals are necessary to complete the integration space of the variables for
$F_3(x)$, and they give the correct limits for $x\rightarrow \pm\infty$ (in a first
 calculation the last four were lost and the limit for $x$ going to $\infty$ was wrong). It is
required a transformation of the integrals on a triangular region with the Fubini's theorem.

The probability distribution is obtained with a derivative of eq.~\ref{eq:equation_50} in the $x$
variable.

\begin{equation}\label{eq:equation_60}
\begin{aligned}
P_{xg2}(x)=&\frac{1}{x^2}\Big [\int_0^\infty\mathrm{d}\beta P_3(\beta)\int_\beta^\infty\mathrm{d}\xi P_1(\xi)\xi
P_2(\xi\frac{1-x}{x})-\int_{-\infty}^0\mathrm{d}\beta P_3(\beta)\int_\beta^0\mathrm{d}\xi P_1(\xi)\xi
P_2(\xi\frac{1-x}{x})+\\
&\int_0^\infty\mathrm{d}\xi P_1(\xi)\int_\xi^\infty\mathrm{d}\beta P_3(\beta)\beta P_2(\beta\frac{-1-x}{x})
-\int_{-\infty}^0\mathrm{d}\xi P_1(\xi)\int_\xi^0\mathrm{d}\beta P_3(\beta)\beta P_2(\beta\frac{-1-x}{x})+\\
&\int_{-\infty}^0\mathrm{d}\beta P_3(\beta)\int_0^\infty\mathrm{d}\xi P_1(\xi)\xi P_2(\xi\frac{1-x}{x})+
\int_{-\infty}^0\mathrm{d}\xi P_1(\xi)\int_0^\infty\mathrm{d}\beta P_3(\beta)\beta P_2(\beta\frac{-1-x}{x})\Big]\\
\end{aligned}
\end{equation}

The first four integrals can be rearranged with the Fubini's theorem and summed with the last two
integrals giving:
\begin{equation}\label{eq:equation_70}
\begin{aligned}
P_{xg_2}(x)=&\frac{1}{x^2}\Big[\int_{0}^\infty\mathrm{d}\xi P_1(\xi)\xi
P_2(\xi\frac{1-x}{x})\int_{-\infty}^\xi\mathrm{d}\beta P_3(\beta)-\\
&\int_{-\infty}^0\mathrm{d}\xi P_1(\xi)\xi
P_2(\xi\frac{1-x}{x})\int_{-\infty}^\xi\mathrm{d}\beta P_3(\beta)+\\
&\int_0^\infty\mathrm{d}\beta P_3(\beta)\beta P_2(\beta\frac{-1-x}{x})\int_{-\infty}^\beta \mathrm{d}\xi P_1(\xi)-\\
&\int_{-\infty}^0\mathrm{d}\beta P_3(\beta)\beta P_2(\beta\frac{-1-x}{x})\int_{-\infty}^\beta \mathrm{d}\xi
P_1(\xi)\Big]
\end{aligned}
\end{equation}
that can be recast in:
\begin{equation}\label{eq:equation_70a}
\begin{aligned}
P_{xg_2}(x)=&\frac{1}{x^2}\Big[\int_{-\infty}^\infty\mathrm{d}\xi \, \big|\xi\big|\, P_1(\xi)
P_2(\xi\frac{1-x}{x})\int_{-\infty}^\xi\mathrm{d}\beta P_3(\beta)+\\
&\int_{-\infty}^\infty\mathrm{d}\beta\, \big|\beta\big|\, P_3(\beta)
P_2(\beta\frac{-1-x}{x})\int_{-\infty}^\beta\,  \mathrm{d}\xi P_1(\xi)\Big]\\
\end{aligned}
\end{equation}

This last form is easer to handle numerically or analytically. MATHEMATICA does not give an
analytical form of the integrals with Gaussian probability distributions,
$\mathrm{erf}$-functions do not allow an analytical result.
The numerical integrals have convergence
problems around $x\approx 0$ even if the results are excellent. The two separate regions of
probability for $x$ are perfect. The function requires three energies and three standard deviations
to produce the results. The three energies should be the unperturbed ones but it is evident that
these are impossible to have.

\section{Three strip COG probability distribution}

Around $x_{g2}=0$ we see an anomaly that produces an incorrect reconstruction of the impact point.
In this region we can try to use the three strip COG algorithm. It has no singularity around
$x_{g3}=0$ (it has singularities around $x_{g3}=\pm 0.5$~\cite{landi01}). Even if the noise is
higher than the $x_{g2}$ it could be convenient to test this strategy in a best fit.
For this task it is required the probability distribution.

As for $x_{g2}$ we have to work in a three dimensional space.

\[a_1=\xi,\ \  a_2=\eta,\ \  a_3=\beta\ \ \ \ x_{g3}=\frac{\xi-\beta}{\xi+\eta+\beta} \]

\begin{figure} [h!]
\begin{center}
\includegraphics[scale=0.7]{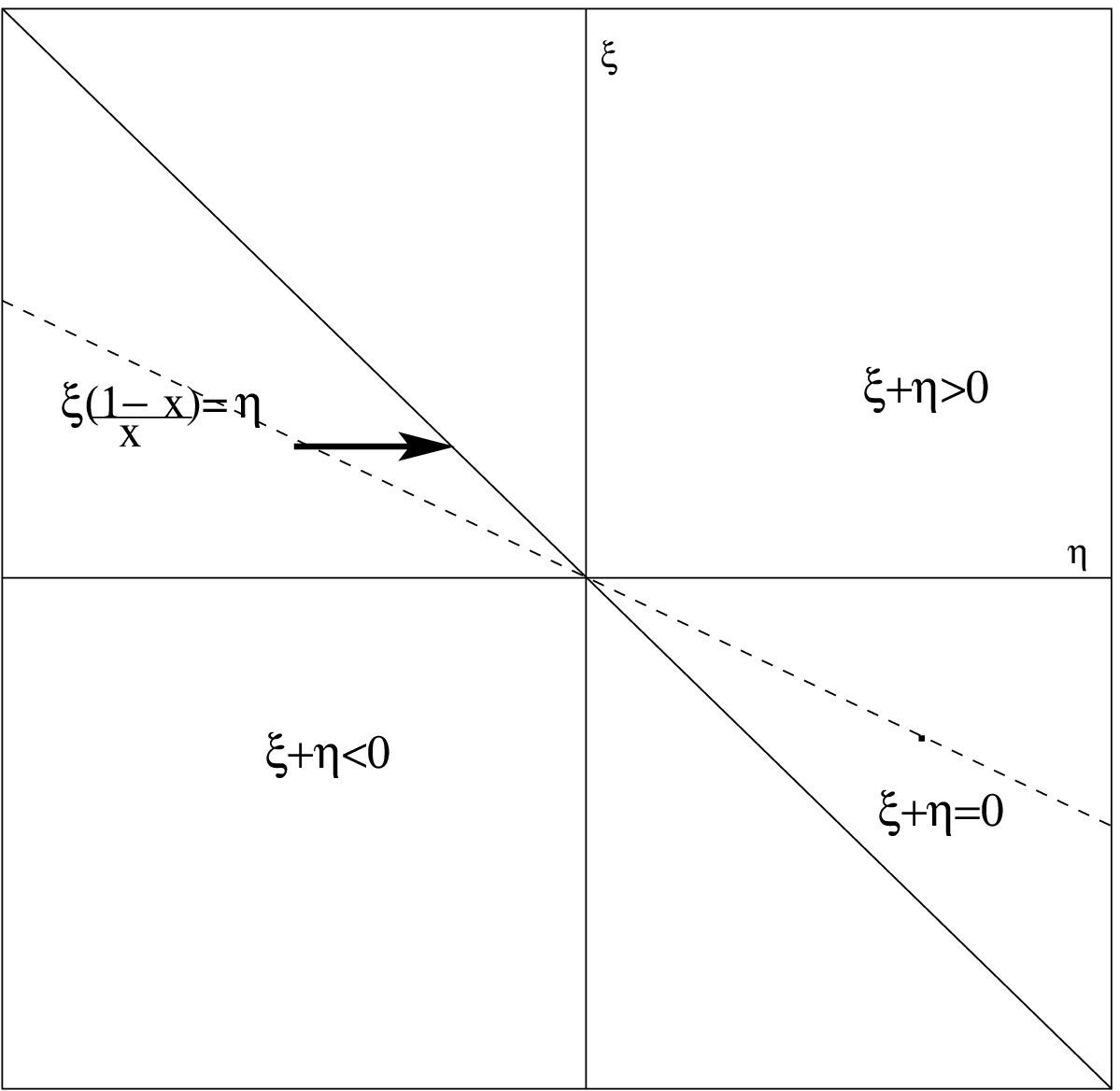}
\caption{\em Sector of the plane $(\eta,\xi)$ where $\xi+\eta>0$ or
$\xi+\eta<0$ and its boundary $\xi+\eta=0$, the dashed line is the line $\xi(1-x)/x=\eta$
for negative $x$, the arrow indicates the $\eta$ integration-path for $(\xi+\eta)<0$
}\label{fig:figure_10}
\end{center}
\end{figure}

\begin{figure} [h!]
\begin{center}
\includegraphics[scale=0.6]{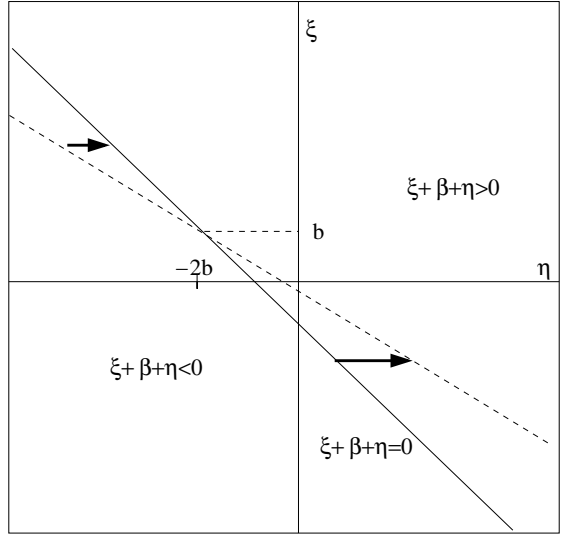}
\caption{\em Sector of the plane $(\eta,\xi)$ where $\xi+\eta+\beta>0$ or
$\xi+\eta+\beta<0$ and its boundary $\xi+\eta+\beta=0$, the dashed line has the equation
 $\xi\,(1-x)+b\,(-1-x)-\eta\,x=0$
for negative $x$, the arrows indicate the $\eta$ integration-paths for $(\xi+\eta+\beta)<0$
and for $(\xi+\eta+\beta)>0$
}\label{fig:figure_11}
\end{center}
\end{figure}

For $\beta=0$ and $\xi+\eta>0$ we have the plot of fig.~\ref{fig:figure_10} (identical to fig.~\ref{fig:figure_3})
but now we have to explore the condition:
\begin{equation}
    \frac{\xi-\beta}{\xi+\eta+\beta}<x.
\end{equation}
We have two possibilities $\xi+\eta+\beta>0$ and $\xi+\eta+\beta<0$, in any case we have
two limiting planes:
\[ \xi(1-x)+\beta(-1-x)-\eta\, x=0\ \ \ \ \ \mathrm{and}\ \ \ \ \  \xi+\eta+\beta=0 \]
for $x<0$ and $\beta=0$, the traces of these two planes in the plane $\eta,\xi$ are those of
fig.~\ref{fig:figure_10}. Let us see the case with $\beta>0$ and $\beta=b$. The traces of
the two planes are those of fig.~\ref{fig:figure_11}. The intersection point is $\{-2\,b,b\}$
and it changes with $\beta$. It is evident that the only change in moving from $\beta>0$ and
$\beta<0$ is the intersection point that changes its sign, and each line of the plot moves parallel
to itself.
Than $F_1^{xg3}(x)$ becomes ($x<0$):
\begin{equation}\label{eq:equation_100}
\begin{aligned}
    F_1^{xg3}(x)=&\int_{-\infty}^\infty\,\mathrm{d}\beta\,P_3(\beta)\int_\beta^\infty\,\mathrm{d}\xi\,P_1(\xi)
    \int_{\xi\frac{(1-x)}{x}+\beta\frac{(-1-x)}{x}}^{-\xi-\beta}P_2(\eta)\mathrm{d}\eta+\\
    &\int_{-\infty}^\infty\,\mathrm{d}\beta\,P_3(\beta)\int_{-\infty}^\beta,\mathrm{d}\xi\,P_1(\xi)
    \int_{-\xi-\beta}^{\xi\frac{(1-x)}{x}+\beta\frac{(-1-x)}{x}}P_2(\eta)\mathrm{d}\eta\\
\end{aligned}
\end{equation}

\begin{figure} [h!]
\begin{center}
\includegraphics[scale=0.6]{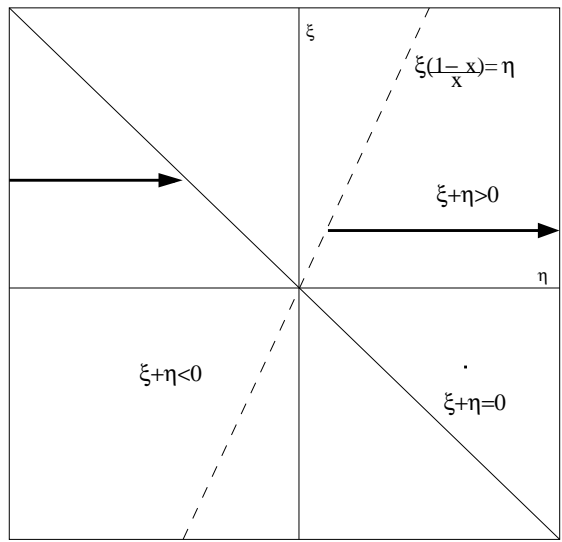}
\caption{\em Sector of the plane $(\eta,\xi)$ where $\xi+\eta>0$ or
$\xi+\eta<0$ and its boundary $\xi+\eta=0$, the dashed line is the line $\xi(1-x)/x=\eta$ for
positive $x$. The integration regions are that with the arrows and must cover positive and negative
values of $\xi$  }\label{fig:figure_12}
\end{center}
\end{figure}

For $x\geq 0$ the traces of the two planes for $\beta=0$ are illustrated in
fig.~\ref{fig:figure_4}, and its identical fig.~\ref{fig:figure_12}.
With $\beta\neq 0$ the traces of the two planes
become these of fig.~\ref{fig:figure_13}. The arrows indicate the integration paths.
The intersection of the two lines are in the point $\{-2b,b\}$ and the $F_2^{xg3}(x)$ becomes:
\begin{figure} [h!]
\begin{center}
\includegraphics[scale=0.6]{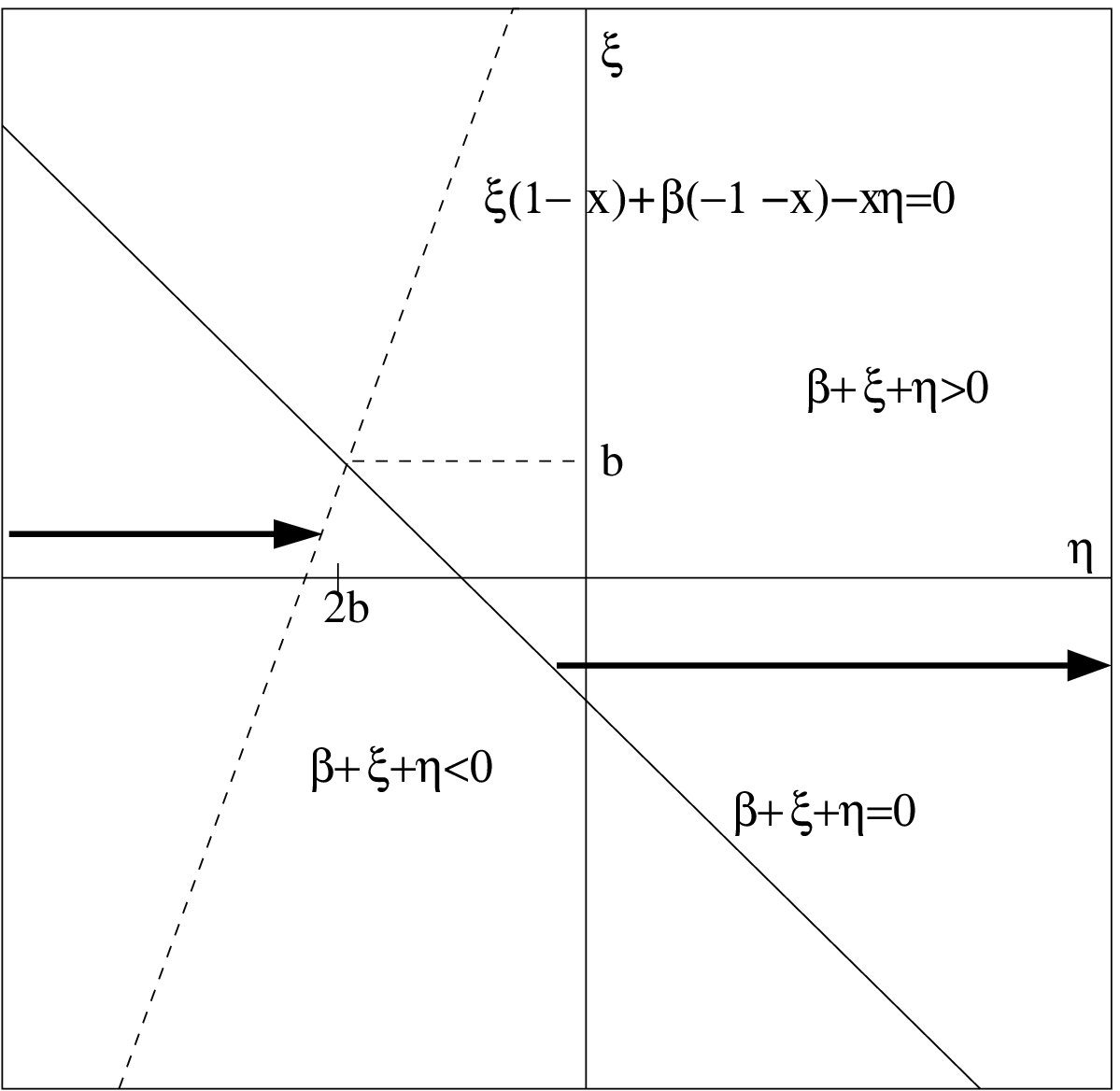}
\caption{\em Sector of the plane $(\eta,\xi)$ where $\xi+\eta+\beta>0$ or
$\xi+\eta+\beta<0$ and its boundary $\xi+\eta+\beta=0$, the dashed line is the line $\xi(1-x)+\beta(-1-x)-x\eta=0$ for
positive $x$. The integration regions are that with the arrows and must cover positive and negative
values of $\xi$  }\label{fig:figure_13}
\end{center}
\end{figure}
\begin{equation}\label{eq:equation_110}
\begin{aligned}
    F_2^{xg3}(x)=&\int_{-\infty}^\infty\,\mathrm{d}\beta\,P_3(\beta)\int_{-\infty}^\beta\,\mathrm{d}\xi\,P_1(\xi)
    \int_{-\xi-\beta}^{\infty}P_2(\eta)\mathrm{d}\eta+\\
    &\int_{-\infty}^\infty\,\mathrm{d}\beta\,P_3(\beta)\int_\beta^{\infty}\,\mathrm{d}\xi\,P_1(\xi)
    \int_{\xi\frac{(1-x)}{x}+\beta\frac{(-1-x)}{x}}^{+\infty}P_2(\eta)\mathrm{d}\eta+\\
    &\int_{-\infty}^\infty\,\mathrm{d}\beta\,P_3(\beta)\int_{-\infty}^\beta,\mathrm{d}\xi\,P_1(\xi)
    \int_{-\infty}^{\xi\frac{(1-x)}{x}+\beta\frac{(-1-x)}{x}}P_2(\eta)\mathrm{d}\eta\\
    &\int_{-\infty}^\infty\,\mathrm{d}\beta\,P_3(\beta)\int_\beta^{+\infty},\mathrm{d}\xi\,P_1(\xi)
    \int_{-\infty}^{-\xi-\beta}P_2(\eta)\mathrm{d}\eta
\end{aligned}
\end{equation}
It is easy to prove that $\lim_{x\rightarrow -\infty}F_1^{xg3}(x)=0$ and
$\lim_{x\rightarrow +\infty}F_2^{xg3}(x)=1$. In fact the first limit is easy given that
$\lim_{x\rightarrow \pm\infty}(1-x)/x=-1$. With this position eq.~\ref{eq:equation_100} has
the limits of the last integrals identical, and the integrals are zero.
For $x\rightarrow +\infty$ the integrals of eq.~\ref{eq:equation_110} become:
\begin{equation}\label{eq:equation_120}
\begin{aligned}
    F_2^{xg3}(+\infty)=&\int_{-\infty}^\infty\,\mathrm{d}\beta\,P_3(\beta)\int_{-\infty}^\beta\,\mathrm{d}\xi\,P_1(\xi)
    \int_{-\xi-\beta}^{\infty}P_2(\eta)\mathrm{d}\eta+\\
    &\int_{-\infty}^\infty\,\mathrm{d}\beta\,P_3(\beta)\int_\beta^{\infty}\,\mathrm{d}\xi\,P_1(\xi)
    \int_{-\xi-\beta}^{+\infty}P_2(\eta)\mathrm{d}\eta+\\
    &\int_{-\infty}^\infty\,\mathrm{d}\beta\,P_3(\beta)\int_{-\infty}^\beta,\mathrm{d}\xi\,P_1(\xi)
    \int_{-\infty}^{-\xi-\beta}P_2(\eta)\mathrm{d}\eta\\
    &\int_{-\infty}^\infty\,\mathrm{d}\beta\,P_3(\beta)\int_\beta^{+\infty},\mathrm{d}\xi\,P_1(\xi)
    \int_{-\infty}^{-\xi-\beta}P_2(\eta)\mathrm{d}\eta
\end{aligned}
\end{equation}
The first and third integrals have identical integration limits in the variables $\beta$ and
$\xi$,  and the sum of the last integrals produces the normalization of the
probability distribution $P_2(\eta)$. Identically for the
second and forth integrals. The two remaining integrals add completing the normalization of the
probability $P_1(\xi)$ that is multiplied by the normalization of $P_3(\beta)$ giving 1.

Now, after this consistency check, we can extract the probability $P_{xg3}(x)$
differentiating $F_1^{xg3}(x)$ and $F_2^{xg3}(x)$ respect to $x$. The result is:
\begin{equation}\label{eq:equation_120}
\begin{aligned}
    P_{xg3}(x)=&\frac{1}{x^2}\Big[\int_{-\infty}^{+\infty}\mathrm{d}\beta\,P_3(\beta)\int_\beta^{+\infty}\mathrm{d}\xi\,
    P_1(\xi)P_2(\xi\frac{1-x}{x}+\beta\frac{-1-x}{x})(-\beta+\xi)+\\
    &\int_{-\infty}^{+\infty}\mathrm{d}\beta\,P_3(\beta)\int_{-\infty}^{\beta}\mathrm{d}\xi\,P_1(\xi)P_2(\xi\frac{1-x}{x}+
    \beta\frac{-1-x}{x})(\beta-\xi)\Big]
\end{aligned}
\end{equation}
With the transformation $\delta=\xi-\beta$, eq.~\ref{eq:equation_120} becomes:
\begin{equation}\label{eq:equation_130}
\begin{aligned}
    P_{xg3}(x)=&\frac{1}{x^2}\Big[\int_{-\infty}^{+\infty}\mathrm{d}\beta\,P_3(\beta)\int_0^{+\infty}\mathrm{d}\delta\,
    P_1(\delta+\beta)P_2(\delta\frac{1-x}{x}-2\beta)\delta-\\
    &\int_{-\infty}^{+\infty}\mathrm{d}\beta\,P_3(\beta)\int_{-\infty}^{0}\mathrm{d}\delta\,P_1(\delta+\beta)P_2(\delta\frac{1-x}{x}
    -2\beta)\delta\Big]
\end{aligned}
\end{equation}
or better:
\begin{equation}\label{eq:equation_130}
    P_{xg3}(x)=\frac{1}{x^2}\Big[\int_{-\infty}^{+\infty}\mathrm{d}\beta\,P_3(\beta)\int_{-\infty}^{+\infty}\mathrm{d}\delta\,
    P_1(\delta+\beta)P_2(\delta\frac{1-x}{x}-2\beta)\big|\delta\big|\Big]
\end{equation}
That can be recast in a form more appropriate for the coming developments:
\begin{equation}\label{eq:equation_130a}
    P_{xg3}(x)=\frac{1}{x^2}\Big[\int_{-\infty}^{+\infty}\mathrm{d}\delta\,\big|\delta\big|\,\int_{-\infty}^{+\infty}\mathrm{d}\beta\,P_3(\beta)
    P_1(\delta+\beta)P_2(\delta\frac{1-x}{x}-2\beta)\Big]
\end{equation}
A set of variable transformations reports this form to be identical to equation 21 of ref.~\cite{landi10}.

\subsection{Expression with Gaussian additive noise and few plots}
With additive Gaussian noise, eq.~\ref{eq:equation_130a} shows the possibility of an analytical
expression of the integrals, in fact the integral on $\beta$ is on product of Gaussian functions
and it will give a Gaussian function. The integral on $\delta$ has itself an analytical
expression.  $P_{xg3}(x)$ assumes the form (with a consistent
help of MATHEMATICA):
\begin{equation}\label{eq:equation_130b}
\begin{aligned}
    &P_{xg3}(x)=\Big\{\exp\big[-(\frac{a_1-a_3}{a_1+a_2+a_3}-x)^2\frac{(a_1+a_2+a_3)^2}
    {2[(1-x)^2\sigma_1^2+x^2\sigma_2^2+(1+x)^2\sigma_3^2]}\big]\\
    &\frac{\Big|a_1[
    x\sigma_2^2+2(1+x)\sigma_3^2]+a_3[2(1-x)\sigma_1^2-x\sigma_2^2]+a_2[(1-x)\sigma_1^2+(1+x)
    \sigma_3^2]\Big|}
    {\sqrt{2\pi}[(1-x)^2\sigma_1^2+x^2\sigma_2^2+(1+x)^2\sigma_3^2]^{3/2}}\Big\}\\
\end{aligned}
\end{equation}
This form is simplified, with the suppression of elements with negligible contributions to the
final result (as far as for the x values of our needs). We have an expression of the type
$A\,\mathrm{erf}(A)$ that is approximated as $|A|$. The differences with the full
integral
are really extremely small.  The normalization of the distributions can be numerically
verified in the range $\approx\pm 3$, beyond these values MATHEMATICA gives wrong a normalization
due to the errors introduced by singularities in the integrands. These singularities are
characteristic of the distributions of ratios of Gaussian random variables as shown in ref.~\cite{gnedenko}.

Another form to write eq.~\ref{eq:equation_130b} consists in the observation that the factors each
$\sigma_i$ $i=1,2,3$ are the COG calculated with the exact energies in the reference system of the
strip $i$. Defining $X_3=(a_1-a_3)/(a_1+a_2+a_3)$, we have:
\begin{equation}
\begin{aligned}
    &a_1[
    x\sigma_2^2+2(1+x)\sigma_3^2]+a_3[2(1-x)\sigma_1^2-x\sigma_2^2]+a_2[(1-x)\sigma_1^2+(1+x)
    \sigma_3^2]=\\
    &(a_1-a_3)x\sigma_2^2+(2a_1+a_2)(1+x)\sigma_3^2+(2a_3+a_2)(1-x)\sigma_1^2=\\
    &(a_1+a_2+a_3)[X_3x\sigma_2^2+(1+X_3)(1+x)\sigma_3^2+(X_3-1)(x-1)\sigma_1^2]\\
\end{aligned}
\end{equation}
With the approximation $X_3\approx x$,
a further simplified form of eq.~\ref{eq:equation_130b} can be obtained , this implies a small
increase of the error (even in the normalization):
\begin{equation}
\begin{aligned}
      &P_{xg3}(x)=\Big\{\exp\big[\!-\!\big(\frac{a_1-a_3}{a_1\!+\!a_2\!+\!a_3\!}-x\!\big)^2\frac{(a_1+a_2+a_3)^2}
    {2(\sigma_1^2(1-x)^2\!+\!x^2\sigma_2^2\!+(1\!+\!x)^2\sigma_3^2)}\big]\\
    &\frac{(a_1+a_2+a_3)}
    {\sqrt{2\pi}\sqrt{(\sigma_1^2(1-x)^2+x^2\sigma_2^2+(1+x)^2\sigma_3^2)}}\Big\}\\
\end{aligned}
\end{equation}
The following figure illustrates the probability reproduction of the data. With
MATLAB~\cite{matlab}, we generate a set of 500000 events all at the identical
noiseless energy. The normalized histograms of $x_{g2}$ and $x_{g3}$ at $\theta=0^o$
and at a noiseless energy of 150 ADC five-strip energy are reported with the
calculated probability distributions. The calculated distributions exactly overlaps
the histograms.

%
%
\begin{figure} [h!]
\begin{center}
\includegraphics[scale=0.8]{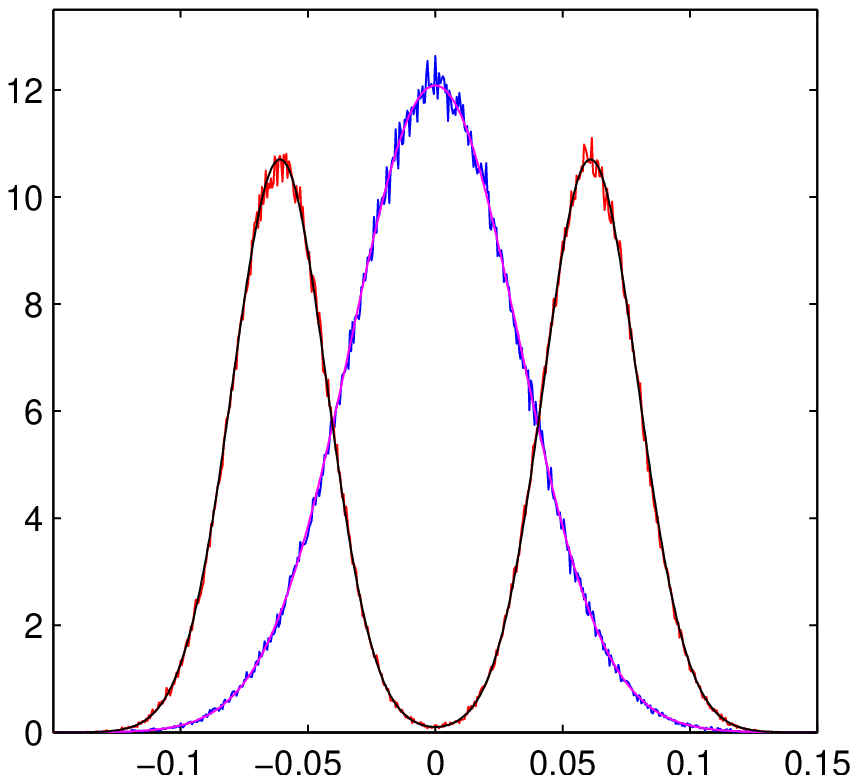}
\includegraphics[scale=0.8]{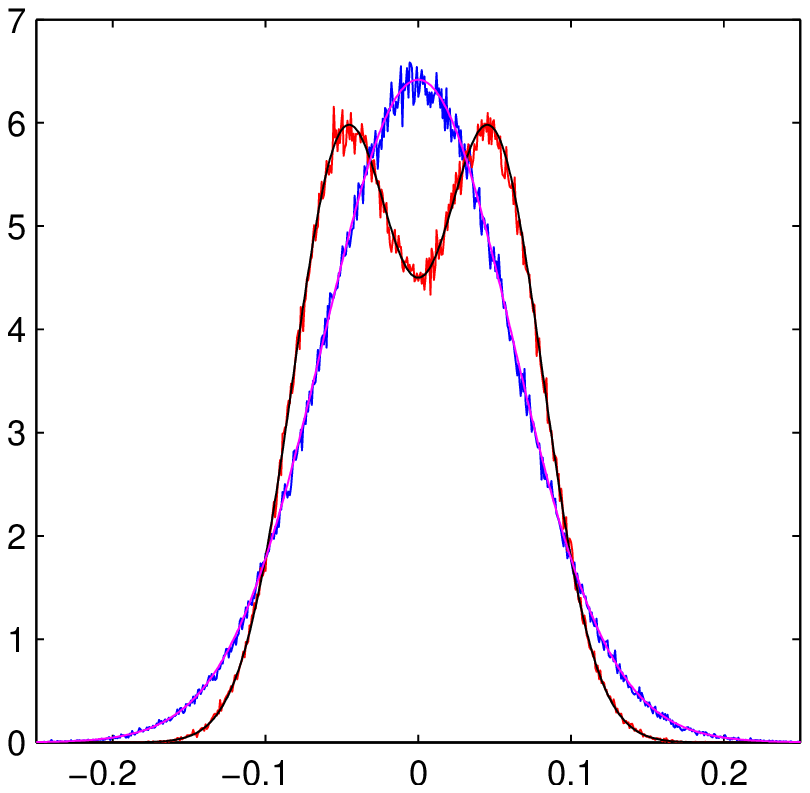}
\caption{\em Right plot. Simulation of the data distributions of $x_{g2}$ in red and $x_{g3}$
blue at the incidence angle $\theta=0^o$,
impact point $\varepsilon=0$ and noiseless energy 150 ADC counts (floating-strip detector). The black curve is the $x_{g2}$ and the
magenta one is the $x_{g3}$ calculated probability distribution. Left plot. Simulations at $\theta= 0^o$,
$\varepsilon=0$ and noiseless energy 150 ADC for normal-strip detector with higher noise ($6.5 ADC$) }\label{fig:figure_15}
\end{center}
\end{figure}

The cumulative distributions for the complete three strip COG (with the gap
at $x \pm 1/2$ as illustrated in ref.~\cite{landi01}) will be reported in a
coming paper.

\section{Conclusions}

The probability distributions for the center-of-gravity  as positioning algorithms are
calculated with the textbook method with the cumulative probability. This method is
very cumbersome but it is reported for completeness. This method was the first, we used
long time ago, for this type of calculations. Other faster methods are reported in
another previous report. Other type of center-of-gravity algorithms will be discussed
in Part II.

\end{document}